\documentclass[a4paper]{jpconf}

\usepackage{graphicx}

\usepackage{amsmath}
\usepackage{amssymb}
\usepackage{amsthm}
\usepackage{paralist}
\usepackage{graphics}
\usepackage{epsfig}
\usepackage[colorlinks=true]{hyperref}
\hypersetup{urlcolor=blue, citecolor=red}
\usepackage{algorithmic}
\usepackage{url}

\newtheorem{theorem}{Theorem}[section]

\begin{document}

\title{Reconstruction and uniqueness of moving obstacles}

\author{Kamen Lozev}

\address{3161 S. Sepulveda Boulevard, Apt. 307, Los Angeles, CA 90034}

\ead{kamen.lozev@gmail.com}

\begin{abstract}
We study the uniqueness and accuracy of the numerical solution of the problem of reconstruction of the shape and trajectory of 
a reflecting obstacle moving in an inhomogeneous medium from travel times, start and end points, and initial angles of ultrasonic
rays reflecting at the obstacle. The speed of sound in the domain when there is no obstacle present is known and provided
as an input parameter which together with the other initial data enables the algorithm to trace ray paths and find their
reflection points. The reflection points determine with high-resolution the shape and trajectory of the obstacle. The method has 
predictable computational complexity and performance and is very efficient when it is parallelized and optimized because only a small 
portion of the domain is reconstructed.
\end{abstract}

\section{Introduction}

Let $\Omega_1(t)$ be a reflecting convex moving obstacle with a smooth boundary in a domain $\Omega_0$  and 
$\overline{\Omega_1(t)} \subset \Omega_0 \subseteq \mathbb{R}^3$. 
Consider an ultrasonic wave, or signal, described by the wave equation 
\begin{equation}\label{waveequation}
u_{tt} - c^2(x,y,z)\Delta{u}=0
\end{equation}

where $c(x,y,z)>0$ is the variable speed of sound in $\Omega(t)=\Omega_0 \backslash  \overline{\Omega_1(t)}$. We consider 
as positive the speed of sound at each point of the domain which is not inside an obstacle and define

$$ c(x,y,z)=0 \textbf{ when } (x,y,z) \in \Omega_1(t) $$ 
and
$$ u |_{\partial{\Omega_1}(t)} = 0 $$

for $t>0$. When $\Omega_1(t)=\emptyset$ then $\Omega(t)=\Omega_0$ and therefore we consider that the speed of sound in $\Omega_0$ is positive
and known when there is no obstacle present. 

We model the signals as rays and look for solutions of the wave equation.
These solutions of the wave equation \cite{E5} are called rays or ray solutions \cite{BB}. We model $\Omega_0$ to be an environment without caustics. 

Let $f(x,y,z)=\frac{1}{c(x,y,z)} > 0$ in $\Omega$. Suppose that for all t we are given all integrals 
$\int_{\gamma} f(l) dl = C_{\gamma}(t)$ where $\gamma$ are broken rays in $\Omega_0$ reflecting at $\partial{\Omega_1(t)}$. 
A broken ray is a ray reflecting at the obstacle and is defined as the union $\gamma=\gamma_1 \bigcup \gamma_2$ of two rays $\gamma_1$ and $\gamma_2$ in
$\Omega_0$ starting at the observation boundary and intersecting at the obstacle's boundary. Then as we know $C_{\gamma}(t)$ correspond to signal travel times in a medium with speed of 
sound $c(x,y,z)$. The shape and trajectory reconstruction problem is to find $\partial{\Omega_1(t)}$ given the sets $C_{\gamma}(t)$ of travel times
where $\gamma \in \Omega_0$, the initial and end positions and take off angles of the rays $\gamma$, and the speed of sound in the whole domain when there is no
obstacle present. 

In our model for the shape and trajectory reconstruction problem rays start at signal transmitters and end at signal receivers with known locations in the observation boundary 
$\partial{\Omega_0}$. In addition, we model the rays to have known initial conditions: the initial zenith and azimuth angles at the transmitter as well as the times
when signals are sent are recorded and are known. Receivers can record the times when signals are received and these times are known as well. 
The combined information from transmitters and receivers provides the data or data points 
$$B_k=(x_{kl}, y_{kl}, z_{kl}, x_{kr}, y_{kr}, z_{kr}, \phi_k, \theta_k, t_k, \xi_k)$$
for the shape and trajectory reconstruction problem where  where $\phi_k$ and $\theta_k$ are the initial incident and azimuth angles of the ray with index k 
from its transmitter, $x_{kl}$, $y_{kl}$, $z_{kl}$ are the coordinates of the transmitter endpoint of the ray, and $x_{kr}$, $y_{kr}$, $z_{kr}$ 
are the coordinates of the receiver endpoint of the ray, $t_k$ is the travel time for the signal and $\xi_k$ is a frequency of the signal.

The ray paths of unbroken rays with known initial conditions are solutions of a system of equations used in the Shooting Method for two-point seismic ray tracing\cite{JG}: 

\begin{align}\label{shooting_method_equations}
\frac{dx}{dt} = c(x,y,z) \sin{\phi}\cos{\theta}\\ 
\frac{dy}{dt} = c(x,y,z) \sin{\phi}\sin{\theta}\\
\frac{dz}{dt} = c(x,y,z) \cos{\phi}\\
\frac{\partial{\phi}}{dt} = -\cos{\phi}( \frac{\partial{c}}{\partial{x}}\cos{\theta} + \frac{\partial{c}}{\partial{y}}\sin{\theta} ) + \frac{\partial{c}}{\partial{z}}\sin{\phi}\\ 
\frac{\partial{\theta}}{dt} = \frac{1}{\sin{\phi}}( \frac{\partial{c}}{\partial{x}}\sin{\theta} - \frac{\partial{c}}{\partial{y}}\cos{\theta} ) 
\end{align}

Systems of equations that present an initial value formulation for the ray equations have origins in acoustics \cite{EL} 
and are used in algorithms for seismic ray tracing \cite{SK, C, BCS}. In other imaging fields, such as non-destructive testing and
biomedical imaging, non-linear ultrasound is studied with a focus on frequency methods\cite{AH,PDNCVBD}. This work provides a mathematical
definition and solution of the shape and trajectory reconstruction problem and is focused on reconstruction and uniques of moving obstacles.  

In the above system of equations (x(t), y(t), z(t)) is the ray position vector, $\phi(t)$ is the incident angle of the ray direction vector with the z axis and $\theta(t)$ is the azimuth angle that the projection of the ray direction vector makes with the positive x axis.

\begin{figure}
\begin{center}
\includegraphics[scale=0.38]{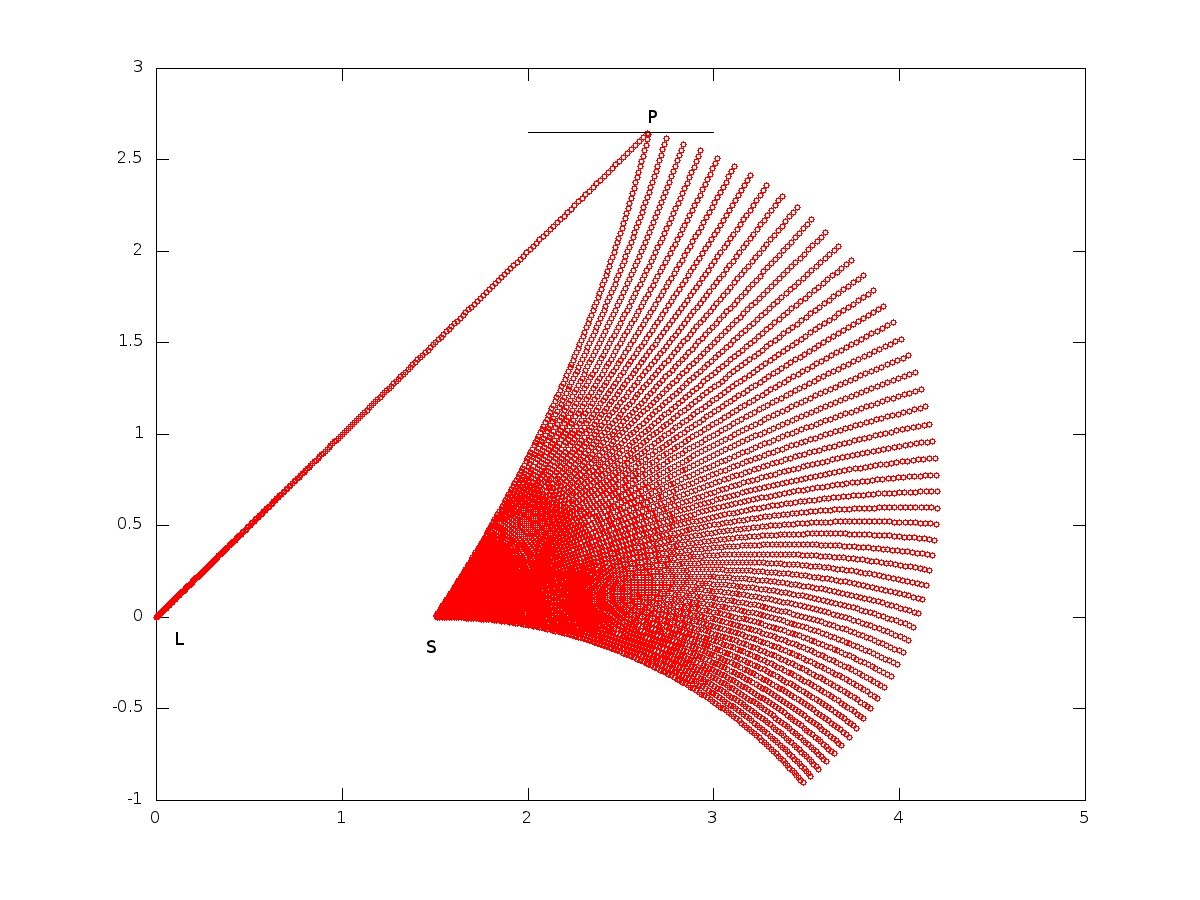} 
\caption{In order to find a solution point P that satisfies the initial conditions for a data point, the algorithm traces a ray from transmitter L 
with initial angles from the data point and traces rays from receiver S with all
possible initial angles. When a receiver ray intersects the transmitter ray at a point P, the algorithm checks whether the sum of
travel times for LP and SP equals the travel time from the data. When the sum of travel times is equal to the travel time from the data point then
P is included in the solution set. This paper describes a new filtering procedure for the solution set which ensures that P is the unique reflection point for 
the measurement ray for the data point.}
\label{algorithm}
\end{center}
\end{figure}

In order to reconstruct the obstacle, we consider the speed of sound $c(x,y,z)$ to be positive and known throughout $\Omega_0$ 
when there is no obstacle present and trace rays from transmitters and receivers as if there is no obstacle. When the sum of ray travel times at an 
intersection point of a transmitter and a receiver ray is equal to the travel time $t_k$ from the corresponding
data point, we infer that the intersection point could be a reflection point from $\partial{\Omega_1}(t)$. 
The algorithm is described in the paper on shape and trajectory reconstruction of moving obstacles\cite{L3} and its operation 
is shown in Figure \ref{algorithm}. 

\clearpage

This work extends the above algorithm
with a filtering phase which ensures that the reconstructed solution is unique and contains only points from $\partial{\Omega_1}(t)$. This
work also extends the first phase of the reconstruction algorithm for computing all reflection points that meet the initial conditions with an 
adaptive computation of the time step for tracing transmitter and receiver branches of broken rays. 
We then analyse the conditions for existence and uniqueness of the solution.

\section{Reconstruction algorithms}\label{variabless_algorithms}

The reconstruction algorithms work in two phases. The first phase finds all points in $\Omega_0$ that are intersection points
of transmitter and receiver rays with initial conditions from the data and for which the sum of travel times of the traced
rays from transmitter and receiver endpoints to the intersection point is equal to the travel time from the corresponding

data point. This section presents the first phase of the algorithm. The next section describes and analyses the second filtering phase.

The input to the following algorithm is the speed of sound $c(x)$ for the domain $\Omega_0$ when there is no obstacle present and 
a set of data points or ray coordinates corresponding to the initial conditions of broken rays and their travel times. 
The output is a set of points in $\mathbb{R}^3$ reconstructed from the input data.

\begin{algorithmic} 
\REQUIRE Set of broken ray data points $B_k=(x_{kl}, y_{kl}, z_{kl}, x_{kr}, y_{kr}, z_{kr}, \phi_k, \theta_k, t_k, \xi_k)$
\REQUIRE Speed of sound $c(x)$ for domain $\Omega_0$ when there is no obstacle  

\COMMENT{Algorithm for Shape and Trajectory Reconstruction of Moving Obstacles}

\COMMENT{Estimated time complexity is $O(T^2A)$ where T is the number of discretization points for the time of flight, 
and A is the number of discretization points for the angle space}           

\FORALL{data points $B_k$}

\STATE{$L=(X_0,Y_0,Z_0)=(x_{kl},y_{kl},z_{kl})$} set this initial position to be position of transmitter
\STATE{$S=(aX_0,aY_0,aZ_0)=(x_{kr},y_{kr},z_{kr})$} set this initial position to be position of receiver 
\STATE{$\Phi_0=\phi_k$}
\STATE{$\Theta_0=\theta_k$}
\STATE{$T_0=0$}
\STATE{$aT_0=0$}

\REPEAT
\STATE
\COMMENT{Compute the next point on the ray from the transmitter by Runge-Kutta step and the ray tracing system \ref{shooting_method_equations}}

\COMMENT{$h$ is a constant value for initializing the time steps}
\STATE{$h_k=h$} 

\REPEAT
\STATE{$X_{s+1} = RK_X(h_k,T_s, X_s,Y_s,Z_s,\Phi_s,\Theta_s$)}
\STATE{$Y_{s+1} = RK_Y(h_k,T_s, X_s,Y_s,Z_s,\Phi_s,\theta_s$)}
\STATE{$Z_{s+1} = RK_Z(h_k,T_s, X_s,Y_s,Z_s,\Phi_s,\Theta_s$)}
\STATE{$\Phi_{s+1} = RK_{\Phi}(h_k,T_s, X_s,Y_s,Z_s,\Phi_s,\Theta_s$)}
\STATE{$\Theta_{s+1} = RK_{\Theta}(h_k,T_s, X_s,Y_s,Z_s,\Phi_s,\Theta_s$)}

\STATE{$X2_{s+1} = RK_X(h_k/2,T_s, X_s,Y_s,Z_s,\Phi_s,\Theta_s$)}
\STATE{$Y2_{s+1} = RK_Y(h_k/2,T_s, X_s,Y_s,Z_s,\Phi_s,\theta_s$)}
\STATE{$Z2_{s+1} = RK_Z(h_k/2,T_s, X_s,Y_s,Z_s,\Phi_s,\Theta_s$)}
\STATE{$\Phi2_{s+1} = RK_{\Phi}(h_k/2,T_s, X_s,Y_s,Z_s,\Phi_s,\Theta_s$)}
\STATE{$\Theta2_{s+1} = RK_{\Theta}(h_k/2,T_s, X_s,Y_s,Z_s,\Phi_s,\Theta_s$)}

\IF{ ($|X_{s+1}-X2_{s+1}|>\epsilon_X$ or $|Y_{s+1}-Y2_{s+1}|>\epsilon_Y$ or $|Z_{s+1}-Z2_{s+1}|>\epsilon_Z$ 
or $|\Phi_{s+1}-\Phi2_{s+1}|>\epsilon_{\Phi}$ or $|\Theta_{s+1}-\Theta2_{s+1}|>\epsilon_{\Theta}$) } 
\STATE $h_k=h_k/2$
\STATE{done=false}
\ELSE
\STATE{done=true}
\ENDIF
\UNTIL{done or ($h_k<h_{min}$)}

\IF{$h_k<h_{min}$}
\STATE
\COMMENT{step size is too small. exit computation for this data point and continue with next data point $B_k$.}
\ENDIF

\STATE{$T_{s+1}=T_s + h_k$}

\IF{$T_{s+1}>t_k$}
\STATE
\COMMENT{We are over the travel time budget $t_k$. Continue with next data point $B_k$.}  
\ENDIF

\STATE $P_{s+1}=(X_{s+1},Y_{s+1}, Z_{s+1})$ point on solution of ray tracing equations with initial values for transmitter that is at time $T_{s+1}$ away from the transmitter L

\IF{!($P_{s+1} \in \Omega_0$)}
\STATE There must be a measurment error. Continue with next data point $B_k$
\ENDIF

\FORALL{initial angles $a\Phi_0, a\Theta_0$ in discretized angle space of the receiver}

\STATE{$\eta_k=h$}

\REPEAT
\STATE 
\COMMENT{Compute the next point on the ray from the receiver by Runge-Kutta step and the ray tracing system \ref{shooting_method_equations}}

\REPEAT
\STATE{$aX_{p+1} = RK_X(\eta_k,aT_p,aX_p,aY_p,aZ_p,a\Phi_p,a\Theta_p$)}
\STATE{$aY_{p+1} = RK_Y(\eta_k,aT_p, aX_p,aY_p,aZ_p,a\Phi_p,a\Theta_p$)}
\STATE{$aZ_{p+1} = RK_Z(\eta_k,aT_p, aX_p,aY_p,aZ_p,a\Phi_p,a\Theta_p$)}
\STATE{$a\Phi_{p+1} = RK_{\Phi}(\eta_k, aT_p, aX_p,aY_p,aZ_p,a\Phi_p,a\Theta_p$)}
\STATE{$a\Theta_{p+1} = RK_{\Theta}(\eta_k, aT_p, aX_p,aY_p,aZ_p,a\Phi_p,a\Theta_p$)}

\STATE{$aX2_{p+1} = RK_X(\eta_k/2,aT_p,aX_p,aY_p,aZ_p,a\Phi_p,a\Theta_p$)}
\STATE{$aY2_{p+1} = RK_Y(\eta_k/2,aT_p, aX_p,aY_p,aZ_p,a\Phi_p,a\Theta_p$)}
\STATE{$aZ2_{p+1} = RK_Z(\eta_k/2,aT_p, aX_p,aY_p,aZ_p,a\Phi_p,a\Theta_p$)}
\STATE{$a\Phi2_{p+1} = RK_{\Phi}(\eta_k/2, aT_p, aX_p,aY_p,aZ_p,a\Phi_p,a\Theta_p$)}
\STATE{$a\Theta2_{p+1} = RK_{\Theta}(\eta_k/2, aT_p, aX_p,aY_p,aZ_p,a\Phi_p,a\Theta_p$)}

\IF{ ($|aX_{s+1}-aX2_{s+1}|>\epsilon_X$ or $|aY_{s+1}-aY2_{s+1}|>\epsilon_Y$ or $|aZ_{s+1}-aZ2_{s+1}|>\epsilon_Z$ 
or $|a\Phi_{s+1}-a\Phi2_{s+1}|>\epsilon_{\Phi}$ or $|a\Theta_{s+1}-a\Theta2_{s+1}|>\epsilon_{\Theta}$) } 
\STATE $\eta_k=\eta_k/2$
\STATE{done=false}
\ELSE
\STATE{done=true}
\ENDIF
\UNTIL{done or ($\eta_k<h_{min}$)}

\IF{$\eta_k<h_{min}$}
\STATE
\COMMENT{step size is too small. exit computation for this data point and continue with next data point.}
\ENDIF

\STATE{$aT_{p+1}=aT_p + \eta_k$}

\STATE $P_{\alpha_{p+1}}=(aX_{p+1},aY_{p+1},aZ_{p+1})$ point on solution of ray tracing equations with initial angles $a\Phi_0$ and $a\Theta_0$ and initial position S, that is time $aT_{p+1}$ away from S

\IF{!($P_{\alpha_{p+1}} \in \Omega_0$)}
\STATE Exit this for loop and continue with next pair of initial angles $a\Phi_0, a\Theta_0$ from outer for loop
\ENDIF

\IF{$distance(P_{s+1},P_{\alpha_{p+1}})<\epsilon_1$ and $|T_{s+1}+aT_{p+1}-t_k|<\epsilon_2$}
\STATE $P=P_k=P_{s+1}$
\COMMENT{Solution for current data point $B_k$ found. Add $(B_k, P)$ to list of all solutions for all data points. Continue with the outer
transmitter loop to look for more solutions for $B_k$.}
\ENDIF

\IF{$T_{s+1}+aT_{p+1}>t_k+\epsilon_2$}
\STATE
\COMMENT{We are over the travel time budget $t_k$. Continue looking for a solution with the next set of initial angles $a\Phi_0, a\Theta_0$.}  
\ENDIF

\UNTIL{threshold for maximum number of time steps from receiver}

\ENDFOR

\UNTIL{threshold for maximum number of time steps from transmitter}

\STATE
\COMMENT{No solution found for $B_k$ due to measurement or other errors. Continue with next data point.}

\ENDFOR

\end{algorithmic}
 
\begin{figure}
\begin{center}
\includegraphics[scale=0.38]{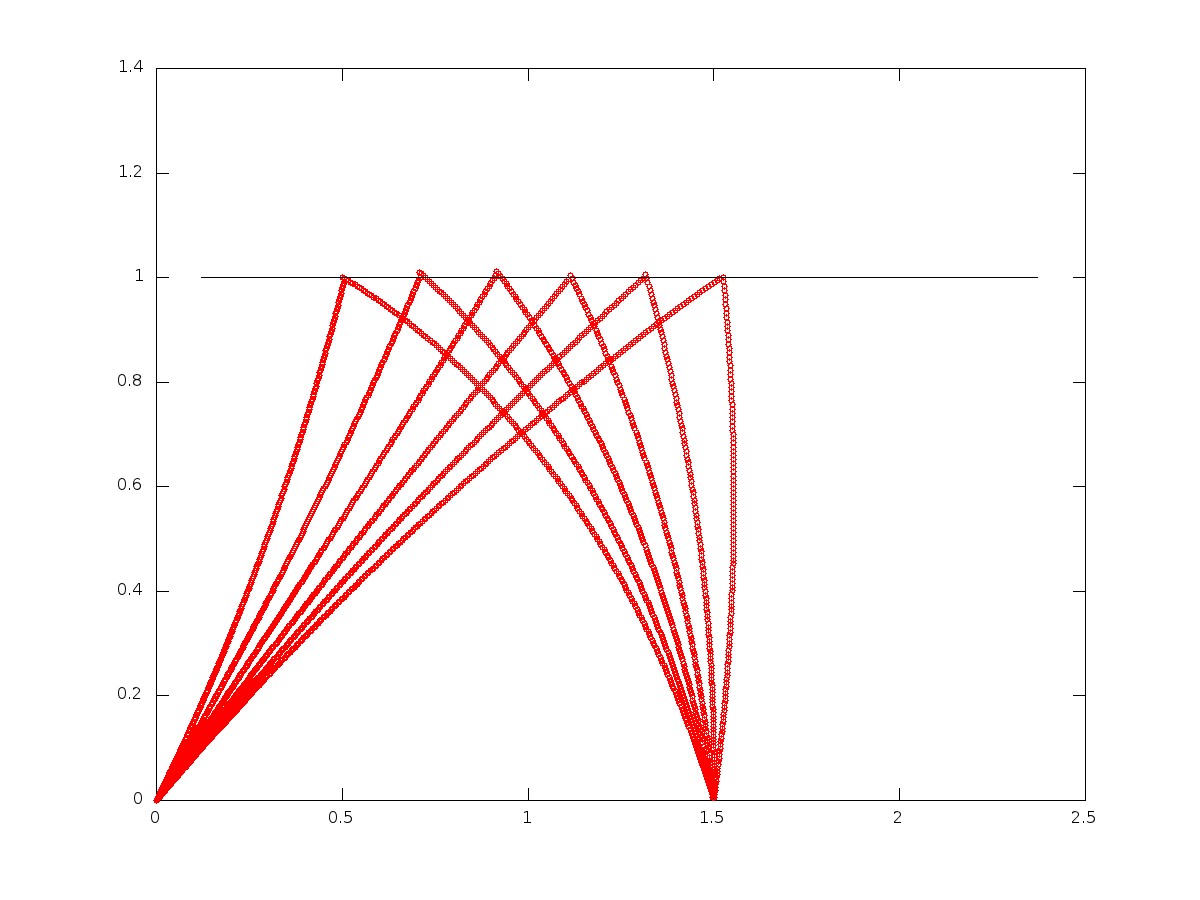} 
\caption{Reconstruction of a line segment in an environment with speed of sound $v(x,y,z)=x+y+1$. Both branches of each broken ray are curves.}
\label{line1}
\end{center}
\end{figure}

\begin{figure}
\begin{center}
\includegraphics[scale=0.38]{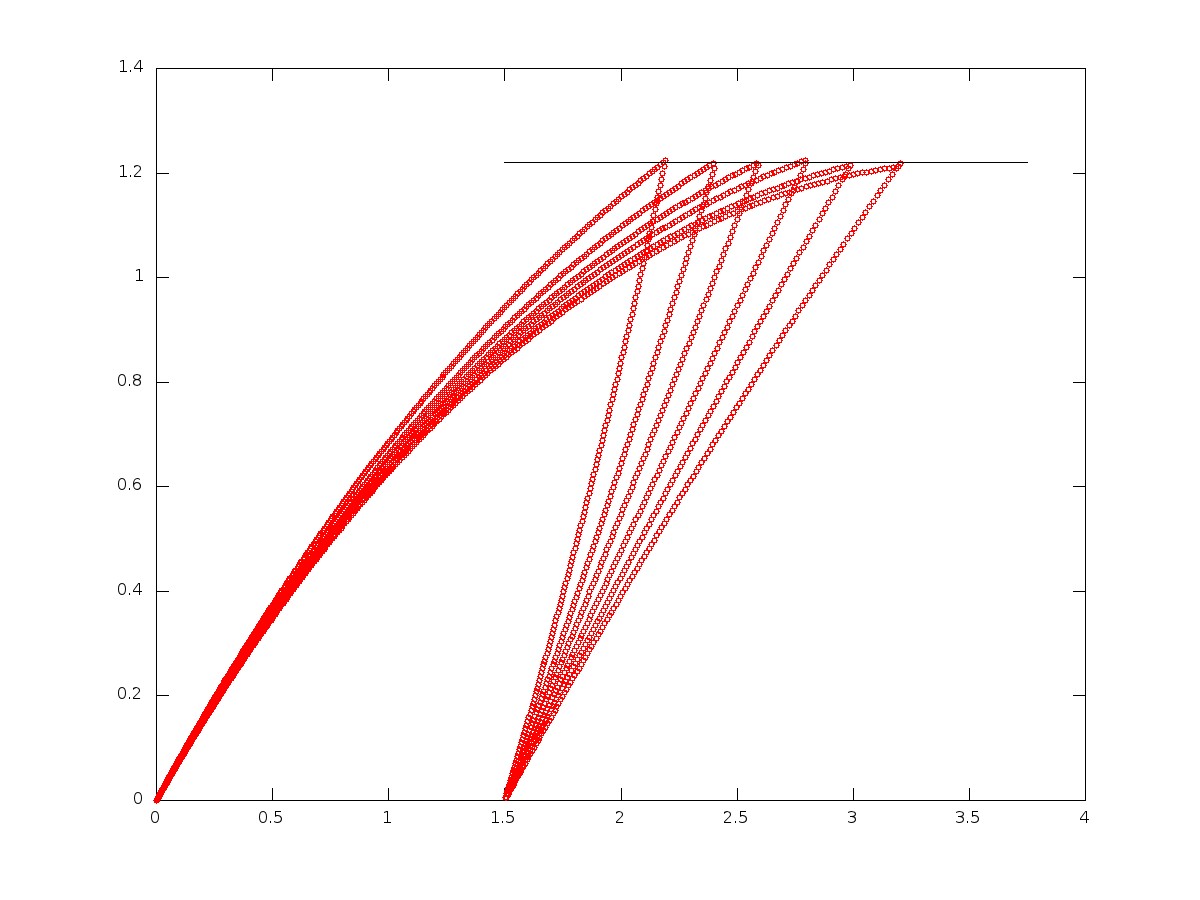} 
\caption{Reconstruction of the same line segment as in Figure \ref{line1} when moving to a new location in the same environment with speed of sound $v(x,y,z)=x+y+1$. Both branches of each broken ray are curves.}
\label{line2}
\end{center}
\end{figure}

\begin{figure}
\begin{center}
\includegraphics[scale=0.38]{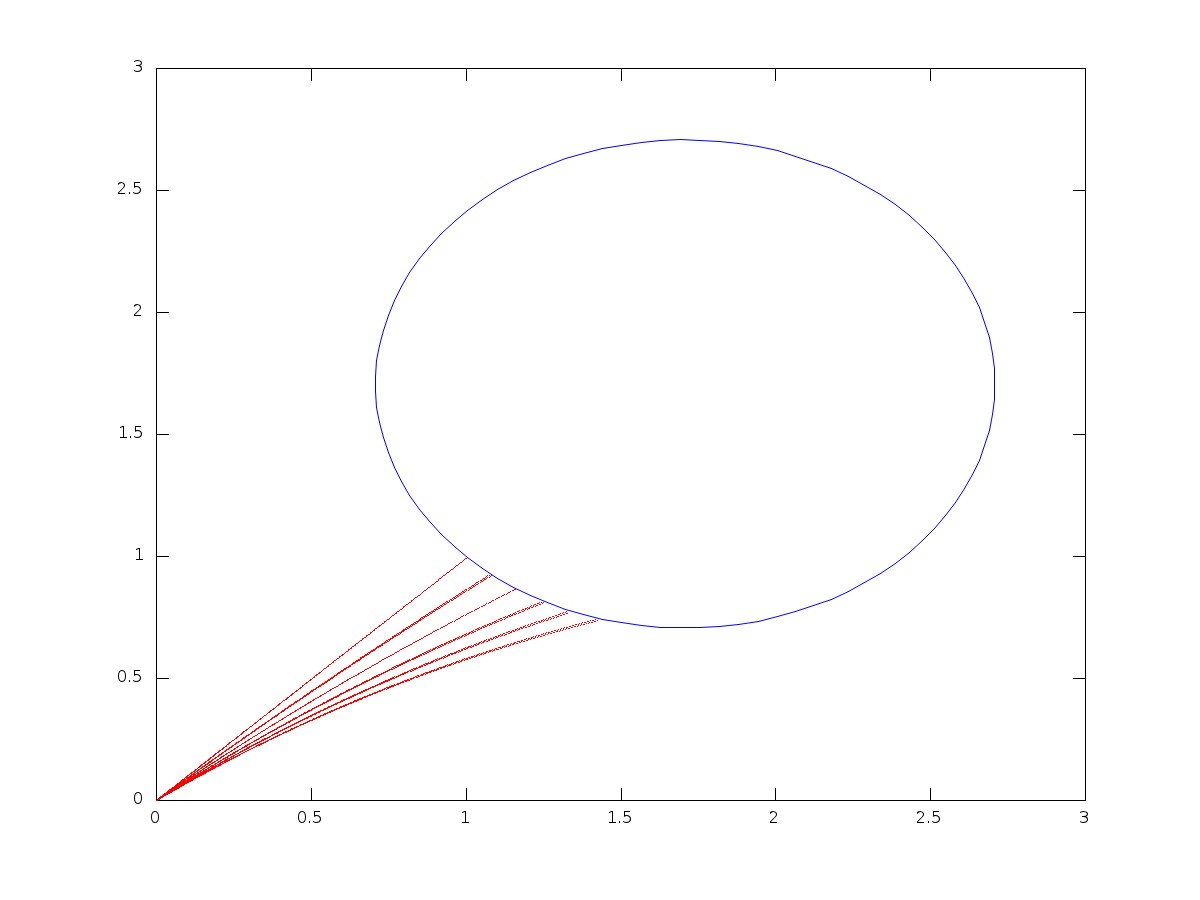} 
\caption{Reconstruction of points from the circular obstacle with a transmitter and receiver located at the origin in the same environment with speed of sound $v(x,y,z)=x+y+1$. 
The obstacle's reflectance is Lambertian and the detected reflected ray travels to the receiver along the same path as the incident ray.}
\label{circle}
\end{center}
\end{figure}

The algorithm uses a Runge-Kutta method for the RK step and is flexible to work with other time-dependent numerical methods. 
When the algorithm is parallelized and caching and other optimization techniques are used then its computational complexity 
is $O(T)$ where T is the number of discretization points for the travel time of a broken ray. For input $\{B_k\}$ from one sampling 
time interval $T_k$,  the algorithm reconstructs the shape of the obstacle during this sampling interval and the trajectory of the obstacle 
is reconstructed when the algorithm is run on the data points for each of the sampling intervals. Resolution of the reconstruction can be very
high because the reconstruction method allows collection and processing of a large number of data points corresponding to different 
points from $\partial{\Omega_1(t)}$. Reconstruction with high resolution by the above algorithm of a neighborhood of points from a moving 
obstacle is shown in Figures \ref{line1}, \ref{line2}, \ref{circle}, \ref{circle2} and \ref{circle3}. 

\clearpage 

\clearpage 

\section{Existence and uniqueness of a reconstructed point}\label{uniqueness}

For a transmitter at $L$ and receiver at $S$ the travel time between L and S along a broken ray $\gamma$ with segments $\gamma_1$ and $\gamma_2$ and
reflection point $P$ is

\begin{equation}
T(L,S)=\int_{\gamma}\frac{ds}{c(s)}=\int_{\gamma_1}\frac{ds}{c(s)} + \int_{\gamma_2}\frac{ds}{c(s)} = t
\end{equation}

For a fixed $t$ and constant speed of sound c the above equation implies that the set of points P is an ellipsoid with focci $L$ and $S$.
This can be seen by multiplying both sides of the equation by $c$ which leads to 

\begin{equation} 
|\gamma_1| + |\gamma_2| = LP + PS=tc=const 
\end{equation}

which is the equation of an ellipsoid with focci $L$ and $S$. 

Therefore, for constant speed of sound $c$ and a data point with transmitter $L$, receiver $S$, initial angles $\phi$ and $\theta$ and
travel time t, a unique point P can be reconstructed because a ray from L with initial angles $\phi$ and $\theta$ travelling along a 
straight line, because of the constant speed of sound, will intersect the ellipsoid in exactly one point P.

For variable speed of sound $c(x,y,z)$ the set of points $P$ is a surface $K$ which is not necessarily convex. In this case a ray from L with 
given initial conditions and travelling along a curve could intersect the surface $K$ in more than one point. In other words there
can be several rays from the receiver that intersect the transmitter ray at different points such that for each intersection point
the sum of travel times from transmitter and receiver to the intersection point is equal to the total travel time. 

In order to find the unique reflection point for the measurement ray, the above algorithm is extended by adding a second filtering
phase after the first phase of the algorithm. The first phase finds all solution points for which the sum of travel times from 
transmitter and receiver to the solution point is equal to the travel time from the data point.  The filtering phase finds the 
unique reflection point by reconstructing 
the shape via several pairs of transmitters and receivers and selecting those points which have been recontructed by a sufficiently
large threshold number $q$, $q \geq 3$, of (transmitter, receiver) pairs. This implies that the data must contain at least
q data points with different transmitter receiver pairs for each reconstructed point. Therefore, the observation boundary must
contain a sufficient number of transmitters and receivers that are located so that each point in $\partial{\Omega_1}(t)$ is seen
from at least q transmitter receiver pairs. The algorithm with the combined first and second phases is as follows.

\begin{algorithmic} 

\REQUIRE Set of broken ray data points $B_k=(x_{kl}, y_{kl}, z_{kl}, x_{kr}, y_{kr}, z_{kr}, \phi_k, \theta_k, t_k, \xi_k)$
\REQUIRE Speed of sound $c(x)$ for domain $\Omega_0$ when there is no obstacle  

\COMMENT{Algorithm for Shape and Trajectory Reconstruction of Moving Obstacles}

\COMMENT{Estimated time complexity is $O(T^2A+N^2)$ where T is the number of discretization points for the time of flight, 
and A is the number of discretization points for the angle space and N is the number of data points. It is possible to
implement the filtering phase with computational complexity $O(N)$ which leads to computational complexity of $O(T^2A+N)$ for the whole algorithm.}           

\FORALL{data points $B_k$}

\STATE{$L=(X_0,Y_0,Z_0)=(x_{kl},y_{kl},z_{kl})$} set this initial position to be position of transmitter
\STATE{$S=(aX_0,aY_0,aZ_0)=(x_{kr},y_{kr},z_{kr})$} set this initial position to be position of receiver 
\STATE{$\Phi_0=\phi_k$}
\STATE{$\Theta_0=\theta_k$}
\STATE{$T_0=0$}
\STATE{$aT_0=0$}

\REPEAT
\STATE
\COMMENT{Compute the next point on the ray from the transmitter by Runge-Kutta step and the ray tracing system \ref{shooting_method_equations}}

\COMMENT{$h$ is a constant value for initializing the time steps}
\STATE{$h_k=h$} 

\REPEAT
\STATE{$X_{s+1} = RK_X(h_k,T_s, X_s,Y_s,Z_s,\Phi_s,\Theta_s$)}
\STATE{$Y_{s+1} = RK_Y(h_k,T_s, X_s,Y_s,Z_s,\Phi_s,\theta_s$)}
\STATE{$Z_{s+1} = RK_Z(h_k,T_s, X_s,Y_s,Z_s,\Phi_s,\Theta_s$)}
\STATE{$\Phi_{s+1} = RK_{\Phi}(h_k,T_s, X_s,Y_s,Z_s,\Phi_s,\Theta_s$)}
\STATE{$\Theta_{s+1} = RK_{\Theta}(h_k,T_s, X_s,Y_s,Z_s,\Phi_s,\Theta_s$)}

\STATE{$X2_{s+1} = RK_X(h_k/2,T_s, X_s,Y_s,Z_s,\Phi_s,\Theta_s$)}
\STATE{$Y2_{s+1} = RK_Y(h_k/2,T_s, X_s,Y_s,Z_s,\Phi_s,\theta_s$)}
\STATE{$Z2_{s+1} = RK_Z(h_k/2,T_s, X_s,Y_s,Z_s,\Phi_s,\Theta_s$)}
\STATE{$\Phi2_{s+1} = RK_{\Phi}(h_k/2,T_s, X_s,Y_s,Z_s,\Phi_s,\Theta_s$)}
\STATE{$\Theta2_{s+1} = RK_{\Theta}(h_k/2,T_s, X_s,Y_s,Z_s,\Phi_s,\Theta_s$)}

\IF{ ($|X_{s+1}-X2_{s+1}|>\epsilon_X$ or $|Y_{s+1}-Y2_{s+1}|>\epsilon_Y$ or $|Z_{s+1}-Z2_{s+1}|>\epsilon_Z$ 
or $|\Phi_{s+1}-\Phi2_{s+1}|>\epsilon_{\Phi}$ or $|\Theta_{s+1}-\Theta2_{s+1}|>\epsilon_{\Theta}$) } 
\STATE $h_k=h_k/2$
\STATE{done=false}
\ELSE
\STATE{done=true}
\ENDIF
\UNTIL{done or ($h_k<h_{min}$)}

\IF{$h_k<h_{min}$}
\STATE
\COMMENT{step size is too small. exit computation for this data point and continue with next data point $B_k$.}
\ENDIF

\STATE{$T_{s+1}=T_s + h_k$}

\IF{$T_{s+1}>t_k$}
\STATE
\COMMENT{We are over the travel time budget $t_k$. Continue with next data point $B_k$.}  
\ENDIF

\STATE $P_{s+1}=(X_{s+1},Y_{s+1}, Z_{s+1})$ point on solution of ray tracing equations with initial values for transmitter that is at time $T_{s+1}$ away from the transmitter L

\IF{!($P_{s+1} \in \Omega_0$)}
\STATE There must be a measurment error. Continue with next data point $B_k$
\ENDIF

\FORALL{initial angles $a\Phi_0, a\Theta_0$ in discretized angle space of the receiver}

\STATE{$\eta_k=h$}

\REPEAT
\STATE 
\COMMENT{Compute the next point on the ray from the receiver by Runge-Kutta step and the ray tracing system \ref{shooting_method_equations}}

\REPEAT
\STATE{$aX_{p+1} = RK_X(\eta_k,aT_p,aX_p,aY_p,aZ_p,a\Phi_p,a\Theta_p$)}
\STATE{$aY_{p+1} = RK_Y(\eta_k,aT_p, aX_p,aY_p,aZ_p,a\Phi_p,a\Theta_p$)}
\STATE{$aZ_{p+1} = RK_Z(\eta_k,aT_p, aX_p,aY_p,aZ_p,a\Phi_p,a\Theta_p$)}
\STATE{$a\Phi_{p+1} = RK_{\Phi}(\eta_k, aT_p, aX_p,aY_p,aZ_p,a\Phi_p,a\Theta_p$)}
\STATE{$a\Theta_{p+1} = RK_{\Theta}(\eta_k, aT_p, aX_p,aY_p,aZ_p,a\Phi_p,a\Theta_p$)}

\STATE{$aX2_{p+1} = RK_X(\eta_k/2,aT_p,aX_p,aY_p,aZ_p,a\Phi_p,a\Theta_p$)}
\STATE{$aY2_{p+1} = RK_Y(\eta_k/2,aT_p, aX_p,aY_p,aZ_p,a\Phi_p,a\Theta_p$)}
\STATE{$aZ2_{p+1} = RK_Z(\eta_k/2,aT_p, aX_p,aY_p,aZ_p,a\Phi_p,a\Theta_p$)}
\STATE{$a\Phi2_{p+1} = RK_{\Phi}(\eta_k/2, aT_p, aX_p,aY_p,aZ_p,a\Phi_p,a\Theta_p$)}
\STATE{$a\Theta2_{p+1} = RK_{\Theta}(\eta_k/2, aT_p, aX_p,aY_p,aZ_p,a\Phi_p,a\Theta_p$)}

\IF{ ($|aX_{s+1}-aX2_{s+1}|>\epsilon_X$ or $|aY_{s+1}-aY2_{s+1}|>\epsilon_Y$ or $|aZ_{s+1}-aZ2_{s+1}|>\epsilon_Z$ 
or $|a\Phi_{s+1}-a\Phi2_{s+1}|>\epsilon_{\Phi}$ or $|a\Theta_{s+1}-a\Theta2_{s+1}|>\epsilon_{\Theta}$) } 
\STATE $\eta_k=\eta_k/2$
\STATE{done=false}
\ELSE
\STATE{done=true}
\ENDIF
\UNTIL{done or ($\eta_k<h_{min}$)}

\IF{$\eta_k<h_{min}$}
\STATE
\COMMENT{step size is too small. exit computation for this data point and continue with next data point.}
\ENDIF

\STATE{$aT_{p+1}=aT_p + \eta_k$}

\STATE $P_{\alpha_{p+1}}=(aX_{p+1},aY_{p+1},aZ_{p+1})$ point on solution of ray tracing equations with initial angles $a\Phi_0$ and $a\Theta_0$ and initial position S, that is time $aT_{p+1}$ away from S

\IF{!($P_{\alpha_{p+1}} \in \Omega_0$)}
\STATE Exit this for loop and continue with next pair of initial angles $a\Phi_0, a\Theta_0$ from outer for loop
\ENDIF

\IF{$distance(P_{s+1},P_{\alpha_{p+1}})<\epsilon_1$ and $|T_{s+1}+aT_{p+1}-t_k|<\epsilon_2$}
\STATE $P=P_k=P_{s+1}$
\COMMENT{Solution for current data point $B_k$ found. Add $(B_k, P)$ to list of all solutions for all data points. Continue with the outer
transmitter loop to look for more solutions for $B_k$.}
\ENDIF

\IF{$T_{s+1}+aT_{p+1}>t_k+\epsilon_2$}
\STATE
\COMMENT{We are over the travel time budget $t_k$. Continue looking for a solution with the next set of initial angles $a\Phi_0, a\Theta_0$.}  
\ENDIF

\UNTIL{threshold for maximum number of time steps from receiver}

\ENDFOR

\UNTIL{threshold for maximum number of time steps from transmitter}

\STATE
\COMMENT{No solution found for $B_k$ due to measurement or other errors. Continue with next data point.}

\ENDFOR

\STATE
\COMMENT{Filter solution set.} 

\FORALL{solution points P}
\STATE
\COMMENT{count how many times P is reconstructed by different transmitter and receiver pairs by checking the distance between P
and all other points Q in the solution set:} 

\IF{$d(P,Q)<\epsilon_3$ and the data points for P and Q have different (transmitter,receiver) pairs}
\STATE{P.count++}
\STATE{remove Q from solution set because it is already counted} 
\ENDIF

\ENDFOR

\FORALL{solutions P in the filtered solution set}

\IF{$P.count < threshold$}

\STATE{remove P from solution set because it is an intangible solution point and 
not a reflection point i.e. reflection points are reconstructed by a sufficiently large number of transmitter receiver pairs.}

\ENDIF

\ENDFOR

\end{algorithmic}

The proof of the correctness of the filtering phase and uniquess conditions on the input data are as follows.

\begin{theorem}{Uniqueness of a point reconstructed from multipile measurements.}
Each point $P \in \partial{\Omega_1(t)}$ can be reconstructed uniquely when the set of data points $\{B_k\}$ for every sampling interval 
contains at least q measurements of $P$ from q different transmitter receiver pairs where q is a sufficiently large threshold number and $q \geq 3$.  
\end{theorem}

Sketch of proof: For each data point $B_k$ the first phase of the above reconstruction algorithm finds one or more solution points $P_1$,...,$P_m$. 
Only one of these solution points is the unique reflection point P for $B_k$'s measurement ray. The remaining solution points for $B_k$ will be referred to as 
intangible solution points. The conditions of the theorem guarantee that for P there are at least
q data points with different transmitter receiver pairs and this implies that P will be counted at least q times by the algorithm. The probability 
$p_k$ that any one of the intangible solution points is also an intangible solution point for another data point and its measurement ray is 
less than 1 and depends on the discretization of the numerical solution. 
Therefore, the probability that any one of the intangible solution points is counted at least q times by the second phase of the algorithm 
and each of these times it is an intangible solution point for the corresponding data point is less than or equal to ${(p_k)}^{q-1}$. 
For sufficiently large q this probability tends to 0, therefore, with probability one, solution points that are not unique reflection points for at least
one measurement ray will be filtered out by the algorithm. Therefore, the conditions of the theorem guarantee that for each $B_k$ a unique reflection
point P with count greater than or equal to q exists because each point from the obstacle's boundary is measured from at least q different 
transmitter receiver pairs.

\section{Reconstruction tests}

Consider a circular reflecting obstacle with Lambertian reflectance in the plane xy moving away from the origin along the line $x=y$ in a medium 
with variable speed of sound $c(x,y)=x+y+1$. We place a transmitter and a receiver at the origin. In this case, 
the domain $\Omega_0$ is a circle of sufficiently large radius that contains the origin. 
Table \ref{Reconstruction_of_a_point_moving_on_the_line_x_y} shows the computation by the algorithm from section
\ref{variabless_algorithms} of the trajectory of a point on the obstacle on the line $x=y$ corresponding to data with different travel times from different
sampling periods.  

\begin{table}[h]
\begin{center}
\begin{tabular}{|ccccccccccccc|}
\hline
xl & yl & zl & xr & yr & zr & $\phi$  & $\theta$ & T & xp & yp & zp & $\Pi$ \\
  \hline
0.00 & 0.00 & 0.00 & 0.00 & 0.00 & 0.00 & 1.57 & 0.79 & 0.25 & 0.09 & 0.09 & 0.00 & $\pi_1$ \\  
0.00 & 0.00 & 0.00 & 0.00 & 0.00 & 0.00 & 1.57 & 0.79 & 0.5 & 0.21 & 0.21 & 0.00 & $\pi_2$ \\ 
0.00 & 0.00 & 0.00 & 0.00 & 0.00 & 0.00 & 1.57 & 0.79 & 0.75 & 0.35 & 0.35 & 0.00 & $\pi_3$\\ 
0.00 & 0.00 & 0.00 & 0.00 & 0.00 & 0.00 & 1.57 & 0.79 & 1.0 & 0.51 & 0.51 & 0.00 & $\pi_4$\\ 
0.00 & 0.00 & 0.00 & 0.00 & 0.00 & 0.00 & 1.57 & 0.79 & 1.25 & 0.71 & 0.71 & 0.00 & $\pi_5$\\ 
0.00 & 0.00 & 0.00 & 0.00 & 0.00 & 0.00 & 1.57 & 0.79 & 1.5 & 0.94 & 0.94 & 0.00 & $\pi_6$\\ 
0.00 & 0.00 & 0.00 & 0.00 & 0.00 & 0.00 & 1.57 & 0.79 & 1.75 & 1.22 & 1.23 & 0.00 & $\pi_7$\\
0.00 & 0.00 & 0.00 & 0.00 & 0.00 & 0.00 & 1.57 & 0.79 & 2.00 & 1.55 & 1.55 & 0.00 & $\pi_8$ \\ 
  \hline
\end{tabular}
\end{center}
\caption{Reconstruction of a point moving with speed $c(x,y)=x+y+1$ on the line 
$x=y$ for a fixed signal frequency $\xi$. 
The initial transmission angles are $\phi=\frac{\pi}{2}$ and $\theta=\frac{\pi}{4}$. The reconstruction data for each row is from a different sampling period.
\label{Reconstruction_of_a_point_moving_on_the_line_x_y}}
\end{table}

We check whether the computation of the above table by the algorithm from \ref{variabless_algorithms} is correct as follows. 
The time for the ray to reach to obstacle can be computed by the formula 

$$ t= \int_{0}^{X} \frac{ds}{c(s)} = \sqrt{2}\int_{0}^{X} \frac{dx}{x+y+1} = \sqrt{2}\int_{0}^{X} \frac{dx}{2x+1}$$

Therefore, $$X=Y=\frac{e^{\sqrt{2}t}-1}{2}$$

By symmetry, for this particular example, this time t is half of the total travel time T.
Then for a travel time $T=2$, or $t=1$, we compute $X=Y=1.55$. This result matches the corresponding result for xp 
and yp from Table \ref{Reconstruction_of_a_point_moving_on_the_line_x_y} obtained by numerical integration. For
$T=1.75$ the numerical computation gives $xp \neq yp$ while by the above formula $X=Y$. 
In this case, the relative error between the values computed by the algorithm and the value computed by the formula is less than 1 percent.

\begin{figure}
\begin{center}
\includegraphics[scale=0.38]{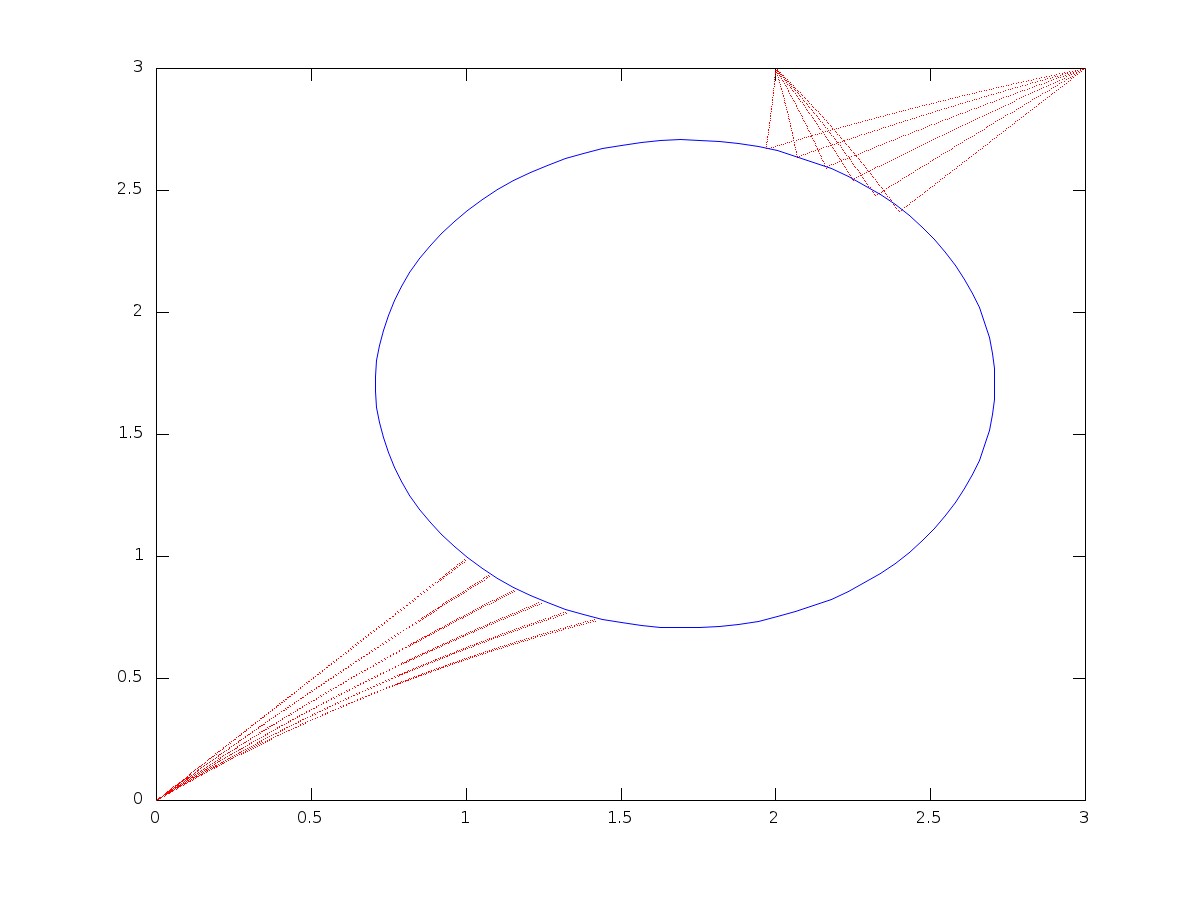} 
\caption{Reconstruction of points from the circular obstacle with a transmitter and receiver located at the origin and another transmitter and receiver located
at $(3,3)$ and $(2,3)$  in the same environment with speed of sound $v(x,y,z)=x+y+1$. 
The obstacle's reflectance is Lambertian and we show the reflected rays intercepting the receiver.}
\label{circle2}
\end{center}
\end{figure}

\begin{figure}
\begin{center}
\includegraphics[scale=0.38]{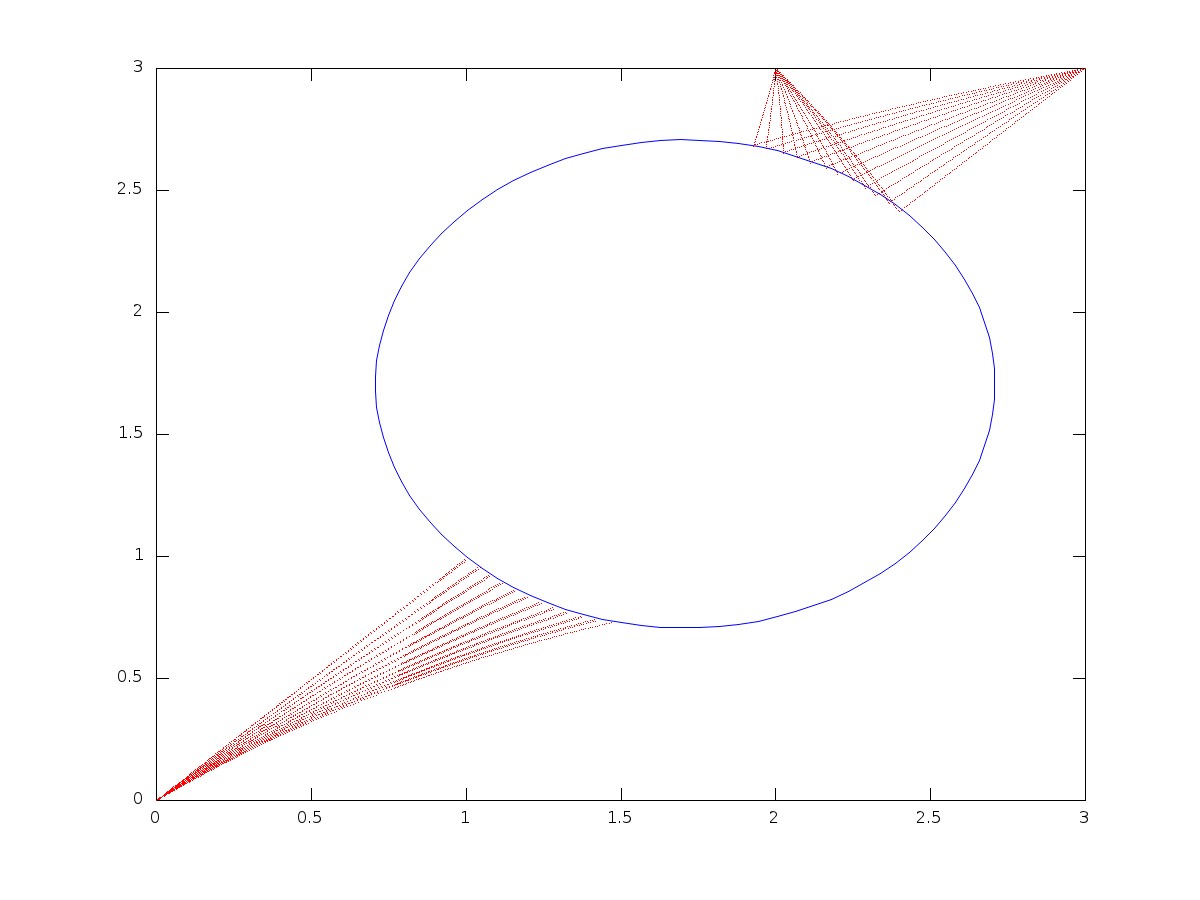} 
\caption{Reconstruction of portions of the boundary of the circular obstacle with a transmitter and receiver located at the origin and another transmitter and receiver 
located at $(3,3)$ and $(2,3)$  in the same environment with speed of sound $v(x,y,z)=x+y+1$. In constrast to Figure \ref{circle2} more rays with initial 
angles that are closer to each other are used for reconstruction and as a result more points are reconstructed that are closer to each other i.e. 
the same portions of the boundary are reconstructed with higher resolution. 
The obstacle's reflectance is Lambertian and we show the reflected rays intercepting the receiver.}
\label{circle3}
\end{center}
\end{figure}

In order to reconstruct more points from $\partial{\Omega_1(t)}$ we can vary the transmission angles. Table \ref{Reconstruction_of_points_on_the_circle_moving_along_x=y}
shows that by varying $\theta$, the initial angle at which we transmit rays from transmitter, we can reconstruct points on the boundary of the circular
obstacle. In contrast, in Table \ref{Reconstruction_of_a_point_moving_on_the_line_x_y} both $\phi$ and $\theta$ are constant. In order to reconstruct
the boundary with higher resolution we can change the initial angles $\theta$ and $\phi$ in smaller steps. Figure \ref{circle2} and Figure \ref{circle3}
show how changing the initial angle $\theta$ at the transmitter in smaller steps leads to higher resolution of the reconstruction.

The speed of the obstacle $v(T)$ must be sufficiently slow compared to the speed of the signals $c(x,y,z)$ during every sampling period so that for the signal 
travel times T from one sampling period, i.e. the times in column T where the sampling period $\Pi$ is the same, for time period $\delta$, where

\begin{equation} 
\delta =  max(T) - min(T) 
\end{equation}

the obstacle does not move by a noticeable amount. $\Pi$ denotes the index or unique id of a sampling time period and $d(\Pi)$ the duration of this time period. 
Therefore, the duration $d(\Pi)$ of the sampling time period $\Pi$, during which rays are sent from the 
transmitter in order to reconstruct the boundary of the obstacle when it is approximately stationary, must be less than or equal to $\delta$ or
\begin{equation} 
d(\Pi) < \delta 
\end{equation}
 
Combining the images of the obstacle from successive sampling intervals reconstructs the shape and trajectory of the moving obstacle. 

\begin{table}[h]
\begin{center}
\begin{tabular}{|ccccccccccccc|}
\hline
xl & yl & zl & xr & yr & zr & $\phi$  & $\theta$ & T & xp & yp & zp & $\Pi$\\
  \hline
0.00 & 0.00 & 0.00 & 0.00 & 0.00 & 0.00 & 1.57 & 0.78 & 1.55 & 1.00 & 0.99 & 0.00 & $\pi_1$\\ 
0.00 & 0.00 & 0.00 & 0.00 & 0.00 & 0.00 & 1.57 & 0.75 & 1.56 & 1.07 & 0.92 & 0.00 & $\pi_1$\\ 
0.00 & 0.00 & 0.00 & 0.00 & 0.00 & 0.00 & 1.57 & 0.71 & 1.58 & 1.15 & 0.86 & 0.00 & $\pi_1$ \\ 
0.00 & 0.00 & 0.00 & 0.00 & 0.00 & 0.00 & 1.57 & 0.68 & 1.61 & 1.24 & 0.81 & 0.00 & $\pi_1$ \\ 
0.00 & 0.00 & 0.00 & 0.00 & 0.00 & 0.00 & 1.57 & 0.66 & 1.65 & 1.32 & 0.77 & 0.00 & $\pi_1$ \\ 
0.00 & 0.00 & 0.00 & 0.00 & 0.00 & 0.00 & 1.57 & 0.64 & 1.70 & 1.42 & 0.74 & 0.00 & $\pi_1$ \\ 
  \hline
\end{tabular}
\end{center}
\caption{Reconstruction of points on the boundary of an obstacle in an environment with speed of sound $c(x,y)=x+y+1$. 
The initial transmission angle $\phi=\frac{\pi}{2}$ and the initial transmission angle $\theta$  varies around $\frac{\pi}{4}$.
The reconstruction data for all rows is from one sampling time period $\pi_1$ during which the obsacle is approximately stationary. 
\label{Reconstruction_of_points_on_the_circle_moving_along_x=y}}
\end{table}

\clearpage 

The reconstruction tests show that for high resolution and performance it is essential that the method is run adaptively. 
In addition to changing the time step, the reconstruction accuracy can be tuned by changing the number of tested angles at the receiver. 

\begin{algorithmic} 
\REQUIRE Set of broken ray data points $B_k=(x_{kl}, y_{kl}, z_{kl}, x_{kr}, y_{kr}, z_{kr}, \phi_k, \theta_k, t_k, \xi_k)$
\REQUIRE Speed of sound $c(x)$ for domain $\Omega_0$ when there is no obstacle  

\COMMENT{Algorithm for Shape and Trajectory Reconstruction of Moving Obstacles}

\COMMENT{Estimated time complexity is $O(T^2A)$ where T is the number of discretization points for the time of flight, 
and A is the number of discretization points for the angle space}           

\FORALL{data points $B_k$}

\STATE{run in parallel the algorithm from section \ref{variabless_algorithms} in order to reconstruct the solution points for data $B_k$}

\IF{no solution found for $B_k$}
\STATE{Run the algorithm from section \ref{variabless_algorithms} with finer discretization of the angle space i.e. double the number of 
tested angles at the receiver in order to reconstruct the point for data $B_k$. Repeat until a solution is found or the angle steps become
smaller than a threshold.}
\ENDIF

\ENDFOR

\STATE{Filter the solution set by the algorithm from section \ref{uniqueness}}

\end{algorithmic}

The above algorithm adapts the angle resolution and via the algorithm from section \ref{variabless_algorithms} it adapts the time step.

\section{Error Analysis}

In floating point arithmetic 
\begin{equation}
t_{LP}+t_{PS}=t+ \epsilon                              
\end{equation}

where $t_{LP}$ is the time for the transmitter segment, $t_{PS}$ is the time for the receiver segment and t the total  
travel time for the broken ray. The check from the algorithm

\begin{algorithmic} 
\IF{$distance(P_{s+1},P_{\alpha_{p+1}})<\epsilon_1$ and $|T_{s+1}+aT_{p+1}-t_k|<\epsilon_2$}
\STATE $P_k=P_{s+1}$
\COMMENT{Solution for current data point $B_k$ found. Continue with next data point $B_{k+1}$}
\ENDIF
\end{algorithmic}

implies that there are many points that are sufficiently close to a solution. The error is determined by the constants
$\epsilon_1$ and $\epsilon_2$ and it is necessary to choose sufficiently small constants for reconstruction with high accuracy and resolution.
Therefore, reconstruction of a point is unique within a ball of sufficiently small diameter which depends on the constants $\epsilon_1$ and $\epsilon_2$.

The error of the solution is also determined by the sum of the two errors from the numerical integrations for the rays from transmitter and
receiver. In addition, the error of the solution is determined by discretization errors. 
The initial angles at the receiver from the numerical solution belong to 
a finite set of initial angles that are tried. The differences between the initial angles $\theta_c, \phi_c$ from the 
numerical solution and real angles $\theta_i, \phi_i$ for the ray at the receiver are 
\begin{equation} 
\Delta_{\theta}=\theta_c - \theta_i 
\end{equation}
and 
\begin{equation}
\Delta_{\phi}=\phi_c - \phi_i  
\end{equation}
  
These differences are guaranteed to be sufficiently small when the initial angles space is tested in sufficiently small equal steps.

\section{Performance Optimizations}

One performance optimization of the algorithm for shape and trajectory reconstruction of moving obstacles 
is the use of a data structure or a database for looking up points on the receiver rays for a given discretization of the initial angles space.
Consider a cover of $\Omega_0$ by a finite number of cubes. Let M be a cube, $\Omega_0 \subset M$, centered at the origin and
with sides parallel to the xy, yz and xz planes. Let $l_m$ be the length of one side of $M$ and divide $M$ into cubes with side length 
$b=\frac{l_m}{N_v}$ where $N_v$ is a natural number that determines the resolution of the mesh. Assign each of the cubes with side b from 
the resulting mesh a unique natural number from 1 to ${N_v}^3$. All precomputed points on receiver rays are stored in a database table RT with the 
following schema:

\begin{table}[h]
\begin{center}
\begin{tabular}{|c|c|c|c|c|c|c|c|}
\hline
pointid & rayid & receiverid & region & t & x & y & z \\
  \hline
\end{tabular}
\end{center}
\caption{Reconstruction database table RT schema: pointid is a unique identifier for each point on a receiver ray,
region is the number of the cube from the mesh that contains the point, t is the time to reach the point from the receiver,
and (x,y,z) the coordinates of the point.
\label{Reconstruction_database}}
\end{table}

The region column corresponds to the number of the cube from the mesh to which point $(x, y, z)$ belongs. The region or cube number of a point can be defined
by the function 

\begin{equation}
m(x,y,z)=\lceil{(\frac{x+ \frac{l_m}{2}}{b})}\rceil \lceil{(\frac{y+ \frac{l_m}{2}}{b})}\rceil \lceil{(\frac{z+ \frac{l_m}{2}}{b})}\rceil
\end{equation}

The optimized first phase of the algorithm with a fixed time step and lookup of cached receiver rays is as follows.

\begin{algorithmic} 
\REQUIRE Set of broken ray data points $B_k=(x_{kl}, y_{kl}, z_{kl}, x_{kr}, y_{kr}, z_{kr}, \phi_k, \theta_k, t_k, \xi_k)$
\REQUIRE Speed of sound $c(x)$ for domain $\Omega_0$ when there is no obstacle.  

\COMMENT{Algorithm for Shape and Trajectory Reconstruction of Moving Obstacles}

\COMMENT{Estimated time complexity is $O(T)$ when using memory databases/data structures of precomputed receiver points.
T is the number of discretization points for the time of flight}           

\FORALL{data points $B_k$}

\STATE{$h_k=\frac{t_k}{N_r}$}
\STATE{$L=(X_0,Y_0,Z_0)=(x_{kl},y_{kl},z_{kl})$} set this initial position to be position of transmitter
\STATE{$S=(aX_0,aY_0,aZ_0)=(x_{kr},y_{kr},z_{kr})$} set this initial position to be position of receiver 
\STATE{$\Phi_0=\phi_k$}
\STATE{$\Theta_0=\theta_k$}
\STATE{$T_0=0$}
\STATE{$aT_0=0$}
\REPEAT
\STATE
\COMMENT{Compute the next point on the ray from the transmitter by Runge-Kutta step and the ray tracing system \ref{shooting_method_equations}}

\STATE{$X_{s+1} = RK_X(h_k,T_s, X_s,Y_s,Z_s,\Phi_s,\Theta_s$)}
\STATE{$Y_{s+1} = RK_Y(h_k,T_s, X_s,Y_s,Z_s,\Phi_s,\theta_s$)}
\STATE{$Z_{s+1} = RK_Z(h_k,T_s, X_s,Y_s,Z_s,\Phi_s,\Theta_s$)}
\STATE{$\Phi_{s+1} = RK_{\Phi}(h_k,T_s, X_s,Y_s,Z_s,\Phi_s,\Theta_s$)}
\STATE{$\Theta_{s+1} = RK_{\Theta}(h_k,T_s, X_s,Y_s,Z_s,\Phi_s,\Theta_s$)}
\STATE{$T_{s+1}=T_s + h_k$}

\STATE $P_{s+1}=(X_{s+1},Y_{s+1}, Z_{s+1})$ point on solution of ray tracing equations with initial values for transmitter that is at time $T_{s+1}$ away from the transmitter L

\IF{!($P_{s+1} \in \Omega_0$)}
\STATE There must be a measurment error. Continue with next data point $B_k$
\ENDIF

\STATE{$region_s=m(X_{s+1}, Y_{s+1}, Z_{s+1})$}

\FORALL{points $P_{\alpha}$ in $region_s$ and adjacent regions}

\IF{$distance(P_{s+1},P_{\alpha})<\epsilon_1$ and $|T_{s+1}+t_{P_{\alpha}}-t_k|<\epsilon_2$}
\STATE $P_k=P_{s+1}$
\COMMENT{Solution for current data point $B_k$ found. Add to solution set for $B_{k}$.}.
\ENDIF

\ENDFOR

\UNTIL{$T_{s+1} \leq t_k$}

\ENDFOR
\end{algorithmic}

The lookup from a relational database for the points in $region_s$ can be implemented via the query

\begin{equation}
P_{\alpha}=\textbf{select from RT (t,x,y,z) where region=m(x,y,z)}
\end{equation}

and an index on region and functional index on the $x, y, z$ columns of table RT. Alternatively, data can be stored in memory.
Computing $region_s=m(x,y,z)$ and looking up the small number of points, close to a constant, 
in a given region and adjacent regions has $O(1)$ computational complexity when the tuples $(t,x,y,z)$ are stored in memory in an array or hash table
where region is the index or key for retrieving tuples from the array or hash table. The computational complexity of the above optimized 
algorithm is then $O(T)$ where $T$ is the number of time steps or discretization points for the travel time. 

Another performance optimization of the reconstruction algorithms is based on the observation that the speed of sound $c(x,y,z)$ is a continuous 
function. This implies that rays with sufficiently close initial conditions reflect at sufficiently close points on the obstacle's boundary and 
conversely that sufficiently close points on the obstacle's boundary could be reconstructed by rays with sufficiently close initial conditions. Once 
the reconstruction algorithm finds the coordinates of one solution point $P_k$ from a given data point $B_k$ then the algorithm optimizes the 
the computation for data points $B_{k+1}$,..., $B_{k+n}$ that have initial conditions that are sufficiently close to $B_k$. For the data points 
$B_{k+1}$,...,$B_{k+n}$ the optimized algorithm starts the search for receiver angles with initial receiver angles that are the same as the receiver 
angles found for $B_k$. As a result, the receiver angles are found with $O(1)$ computational complexity. 
This optimization leads to computational complexity for reconstructing a point of $O(T^2)$. In this section, fixed time steps are used
for brevity and to show the differences with an adaptive time step implementation. In a small neighbourhood of the transmitter angles
$(\phi,\theta)$ of the first data point $B_1$ the optimized first phase of the reconstruction algorithm with a fixed time step is now: 

\begin{algorithmic} 
\REQUIRE Set of broken ray data points $B_k=(x_{kl}, y_{kl}, z_{kl}, x_{kr}, y_{kr}, z_{kr}, \phi_k, \theta_k, t_k, \xi_k)$ with sufficiently close $(\phi,\theta)$ i.e.
$1 \leq k \leq n$, $||(\phi_i, \theta_i),(\phi_j, \theta_j)|| \leq \epsilon_0$, $1 \leq i \leq n$, $1\leq j \leq n$.
\REQUIRE Speed of sound $c(x)$ for domain $\Omega_0$ when there is no obstacle present.  

\COMMENT{Algorithm for Shape and Trajectory Reconstruction of Moving Obstacles}

\COMMENT{Estimated time complexity is $O(T^2)$ where T is the number of discretization points for the time of flight}           

\COMMENT{Reconstruct the coordinates of the first point $B_1$ and associated receiver angles or use an overlapping point from an already reconstructed neighbourhood set}
\STATE{$PX_1, PY_1, PZ_1, R\Phi_1, R\Theta_1$}

\FORALL{data points $B_k$ where $2 \leq k \leq n$}

\STATE{$h_k=\frac{t_k}{N_r}$}
\STATE{$L=(X_0,Y_0,Z_0)=(x_{kl},y_{kl},z_{kl})$} set this initial position to be position of transmitter
\STATE{$S=(aX_0,aY_0,aZ_0)=(x_{kr},y_{kr},z_{kr})$} set this initial position to be position of receiver 
\STATE{$\Phi_0=\phi_k$}
\STATE{$\Theta_0=\theta_k$}
\STATE{$T_0=0$}
\STATE{$aT_0=0$}
\FOR{$s = 0 \to N_r-1$}
\STATE
\COMMENT{Compute the next point on the ray from the transmitter by Runge-Kutta step and the ray tracing system \ref{shooting_method_equations}. $N_r$ is
the number of fixed time steps.}

\STATE{$X_{s+1} = RK_X(h_k,T_s, X_s,Y_s,Z_s,\Phi_s,\Theta_s$)}
\STATE{$Y_{s+1} = RK_Y(h_k,T_s, X_s,Y_s,Z_s,\Phi_s,\Theta_s$)}
\STATE{$Z_{s+1} = RK_Z(h_k,T_s, X_s,Y_s,Z_s,\Phi_s,\Theta_s$)}
\STATE{$\Phi_{s+1} = RK_{\Phi}(h_k,T_s, X_s,Y_s,Z_s,\Phi_s,\Theta_s$)}
\STATE{$\Theta_{s+1} = RK_{\Theta}(h_k,T_s, X_s,Y_s,Z_s,\Phi_s,\Theta_s$)}
\STATE{$T_{s+1}=T_s + h_k$}

\STATE $P_{s+1}=(X_{s+1},Y_{s+1}, Z_{s+1})$ point on solution of ray tracing equations with initial values for transmitter that is at time $T_{s+1}$ away from the transmitter L

\IF{!($P_{s+1} \in \Omega_0$)}
\STATE There must be a measurment error. Continue with next data point $B_k$
\ENDIF

\FORALL{initial angles $a\Phi_0, a\Theta_0$ in discretized angle space in a small neighborhood of $(R\Phi_1, R\Theta_1)$}

\FOR{$p = 0 \to N_r-1$}
\STATE
\COMMENT{Compute the next point on the ray from the receiver by Runge-Kutta step and the ray tracing system \ref{shooting_method_equations}. $N_r$ is
the number of fixed time steps.}
\STATE{$aX_{p+1} = RK_X(h_k,aT_p,aX_p,aY_p,aZ_p,a\Phi_p,a\Theta_p$)}
\STATE{$aY_{p+1} = RK_Y(h_k,aT_p, aX_p,aY_p,aZ_p,a\Phi_p,a\Theta_p$)}
\STATE{$aZ_{p+1} = RK_Z(h_k,aT_p, aX_p,aY_p,aZ_p,a\Phi_p,a\Theta_p$)}
\STATE{$a\Phi_{p+1} = RK_{\Phi}(h_k, aT_p, aX_p,aY_p,aZ_p,a\Phi_p,a\Theta_p$)}
\STATE{$a\Theta_{p+1} = RK_{\Theta}(h_k, aT_p, aX_p,aY_p,aZ_p,a\Phi_p,a\Theta_p$)}
\STATE{$aT_{p+1}=aT_p + h_k$}

\STATE $P_{\alpha_{p+1}}=(aX_{p+1},aY_{p+1},aZ_{p+1})$ point on solution of ray tracing equations with initial angles $a\Phi_0$ and $a\Theta_0$ and initial position the location of the receiver S, 
that is time $aT_{p+1}$ away from S

\IF{!($P_{\alpha_{p+1}} \in \Omega_0$)}
\STATE Exit this for loop and continue with next pair of initial angles $a\Phi_0, a\Theta_0$ from outer for loop
\ENDIF

\IF{$distance(P_{s+1},P_{\alpha_{p+1}})<\epsilon_1$ and $|T_{s+1}+aT_{p+1}-t_k|<\epsilon_2$}
\STATE $P_k=P_{s+1}$
\COMMENT{Solution for current data point $B_k$ found. Add to solution set for $B_k$.}
\ENDIF

\IF{$T_{s+1}+aT_{p+1}>t_k+\epsilon_2$}
\STATE
\COMMENT{We are over the travel time budget $t_k$. Continue looking for a solution with the next set of initial angles $a\Phi_0, a\Theta_0$.}  
\ENDIF

\ENDFOR

\ENDFOR
\ENDFOR

\ENDFOR
\end{algorithmic}

When the reconstruction of the neighborhood $B_k$ is complete, the computed points can be used as the first points or seeds with precomputed
receiver angles for new neighborhoods or patches of the angle space. Thus reconstruction is performed for patches of points which leads to
better performance compared to reconstruction when all points are reconstructed independently.  Reconstruction of each
point within a given patch is performed independently and in parallel with reconstruction of all other points in the patch.

\section{Acknowledgements} I would like to thank Professor Gregory Eskin and Professor James Ralston. I would like to thank the participants in the conferences
WiS\&E 2011, IC-MSQUARE 2012, and WiS\&E 2013.


\section{References}


\begin{thebibliography}{99}

\bibitem{E5}
Eskin G. \emph{Lectures on Linear Partial Differential Equations},
  \emph{Graduate Studies in Mathematics}, vol. 123. American Mathematical
  Society: Providence, Rhode Island, 2011.

\bibitem{BB}
VMBabi$\check{c}$, Buldyrev V. \emph{Short-Wavelength Diffraction Theory}.
  Springer Verlag: Berlin Heidelberg, 1991.

\bibitem{JG}
Julian B, Gubbins D. Three-dimensional seismic ray tracing. \emph{J. Geophys.}
  1977; \textbf{43}:95--1113.

\bibitem{EL}
Eliseevnin V. Analysis of waves propagating in an inhomogeneous medium.
  \emph{Soviet Physics}  1965; \textbf{Acoustics}(10):242--245.

\bibitem{SK}
Sambridge M, Kennet B. Boundary value ray tracing in a heterogeneous medium: a
  simple and versatile algorithm. \emph{Geophys. J. Int.}  1990;
  \textbf{101}:157--168.

\bibitem{C}
{\v C}erven{\'y} V. \emph{Seismic Ray Theory}. Cambridge University Press:
  Cambridge, 2001.

\bibitem{BCS}
Bleinstein N, Cohen J, Stockwell J. \emph{Mathematics of Multidimensional
  Seismic Imaging Migration and Inversion}. Springer Verlag: New York, 2001.

\bibitem{AH}
Arnold W, Hirsekorn S ( (eds.)). \emph{Acoustical Imaging}, vol.~27. Springer
  Netherlands.

\bibitem{PDNCVBD}
Pasovic M, Danilouchkine M, van Neer P, Cachard C, Van Der~Steen AFW, Basset O,
  De~Jong N. Second harmonic inversion for ultrasound contrast harmonic
  imaging. \emph{Physics in Medicine and Biology}  2011; \textbf{56}(11).

\bibitem{L3}
Lozev K. Algorithms for shape and trajectory reconstruction of obstacles in
  domains with variable speed of sound. \emph{J. Physics Conf. Ser.}  2013;
  \textbf{410}(012171):1--4.

\end{thebibliography}
\end{document}